\documentclass[epj,onecolumn]{webofc}
\usepackage[varg]{txfonts}   
\usepackage{url}
\usepackage{color}
\usepackage{epsfig}
\usepackage{graphicx}
\usepackage{subfigure}
\def\ifm#1{\relax\ifmmode#1\else$#1$\fi}
\def\ab{\ifm{\sim}}  
\def\epm{\ifm{e^+e^-}} 

\def\lfigbox#1;#2;{\parbox{#2cm}{\vglue3mm\epsfig{file=#1.eps,width=#2cm}\vglue3mm}}
\def\km{\kern-1.5mm}    \def\kma{\kern-2.5mm}\def\kms{\kern-.75mm}
\def\pt#1,#2,{\ifm{#1\x10^{#2}}}   \def\bye{\end{document}}
\def\strike#1;{\setbox0=\hbox{#1}\rlap{\vrule height3.3pt depth-2.3pt width\wd0}\kern2pt#1\ }

\def\lfigbox#1;#2;{\parbox{#2cm}{\vglue3mm\epsfig{file=#1.eps,width=#2cm}\vglue3mm}}
\def\epm{\ifm{e^+e^-}} 
\def\to{\ifm{\rightarrow}}
 \def\sig{\ifm{\sigma}} 
\def\pic{\ifm{\pi^+\pi^-}}

\newcommand{\ppg}{$\pi\pi\gamma$}
\newcommand{\mmg}{$\mu\mu\gamma$}

\newcommand{\mtrk}{$M_{TRK}$}

\woctitle{KLOE-2 Workshop on $e^+e^-$ collider physics at 1 GeV}
\begin{document}
\title{From Hadronic Cross Section to the measurement of the Vacuum Polarization at KLOE: a fascinating endeavour}
%
%



\author{Graziano Venanzoni\inst{1}\fnsep\thanks{\email{graziano.venanzoni@pi.infn.it}} on behalf of the KLOE-2 Collaboration}

\institute{INFN Sezione di Pisa, Pisa, Italy}

\abstract{%
The KLOE experiment at the $\phi-factory$ DA$\Phi$NE in Frascati 
is the first to have employed Initial State Radiation (ISR)
 to precisely determine the $e^+e^-\to\pi^+\pi^-(\gamma)$ cross section below 1 GeV. Such a measurement is particularly important to test the Standard Model (SM) calculation for the $(g-2)$ of the muon, where a  long standing 3$\sigma$ discrepancy is observed.
I will review the ISR activity in KLOE in the last 18 years from 
the measurement of the hadronic cross section to the first direct determination of the time-like complex running $\alpha(s)$ in the region below 1 GeV.}

%
\maketitle
\section{Introduction: the Endurance expedition}
\begin{figure}[t]
\begin{center}
\mbox{
\includegraphics[width=12.pc]{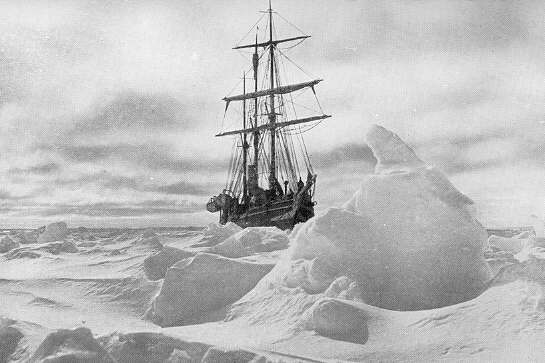}
\hspace*{0.5cm}
\includegraphics[width=11.3pc]{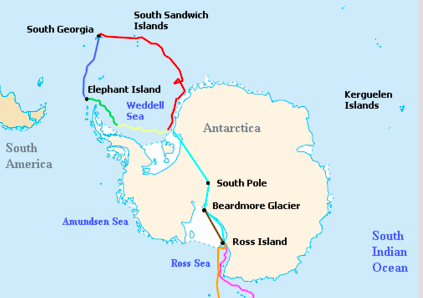}
}
\caption{Left: The Endurance trapped in ice;
{\it —Photo by Frank Hurley, courtesy of the Macklin Collection}; Right: Map of the sea routes of Endurance and James Caird:
in red the voyage of Endurance; in 
 yellow the drift in pack ice; in green the sea-ice drift after Endurance sinks; in blue the voyage of the boat James Caird.}
\label{fig:1} 
\end{center}
\end{figure}

In August 1914, polar explorer Sir Ernest Shackleton boarded the Endurance and set sail for Antarctica, where he planned to cross the last uncharted continent on foot. In January 1915, after battling its way through a thousand miles of pack ice and only a day's sail short of its destination, the Endurance became locked in an island of ice (see Fig~\ref{fig:1}, left). Thus began the legendary ordeal of Shackleton and his crew of twenty-seven men.  For ten months the ice-moored Endurance drifted northwest before it was finally crushed between two ice floes. With no options left,
Shackleton and a 5 members crew left on 24 April 1916 on a 7 m boat, the {\it James Caird}, from 
Elephant Island in the South Shetland Islands
for a near-impossible journey over 850 miles of the South Atlantic's heaviest seas to the island of South Georgia (see Fig.~\ref{fig:1}, right). Shackleton indeed succeeded to reach the island, and on 30 August, after four attempts, Shackleton was able to return to Elephant Island to rescue the party stranded there. Every single man survived.

With due proportion, this epic feat of endurance had many distinctive characters of the modern scientific research: 
experience the raw beauty and test the human spirit in an unknown (and often hostile) territory with perseverance and dedication;
draw conclusions 
from observation, (incomplete) information
and inspiration;
learn from errors.
Such a mental attitude
especially
applies when the object of a research is a precision measurement, as was the trip of the {\it James Caird} boat to South Georgia, a tiny dot of land in a vast and hostile ocean.



\label{intro}

\section{1999-2004: The first KLOE measurement 
of  $\sigma(e^+e^- \rightarrow \pi^+\pi^-)$ with ISR}
After more than twenty years from the first calculations~\cite{Baier1}, the late nineties saw a vigorous and renovated interest towards the use of Initial State Radiation process (ISR in the following) for the precise measurement of the hadronic cross section~\cite{arbuzov1}. The high luminosity reached at flavor factories made it possible to compensate the smaller cross section of the radiative process, making ISR an alternative and competitive method with respect to the traditional energy scan~\cite{tutterev}.
From the theoretical side the development of the PHOKHARA Monte Carlo event generator~\cite{phokhara}, based on exact QED NLO predictions, together with improved Monte Carlo codes for luminosity measurement, like BABAYAGA~\cite{babayaga} or BHWIDE~\cite{bhwide} and MCGPJ~\cite{mcjpg}, made it possible to reach the sub per cent accuracy on the measurement of ISR cross sections~\cite{actis}.

From the experimental side ISR was first performed at KLOE~\cite{spagnolo,cataldi}, to determine the cross section $e^+e^-\to\pi^+\pi^-\gamma$, soon followed by BaBar~\cite{solodov}.
 In the recent years this method was used to study a multitude of final states at BaBar~\cite{anulli}, Belle~\cite{Nakazawa:2015dka,Pakhlova:2008zz} and  BES-III~\cite{Ablikim:2015orh}.  

The first KLOE analysis of $e^+e^-\to\pi^+\pi^-\gamma$ cross section (called KLOE05 in the following) was performed during the years from 1999 to 2004~\cite{Aloisio:2004bu}. 
This analysis was mainly motivated by the discrepancy between the 
Standard Model prediction (SM in the following) and the
measured value of the muon g-2~\cite{Bennett:2006fi}.
A large part of the 
uncertainty on the
theoretical estimates comes from the leading order hadronic
contribution $a_\mu^{\mathrm{HLO}}$, which at
low energies is not calculable by perturbative QCD, but has to be
evaluated with a dispersion integral using measured 
hadronic cross sections.

The process \epm\to\pic\ below $1$ GeV accounts for $\sim 70\%$ of 
$a_{\mu}^{\mathrm{HLO}}$~\cite{Eidelman:1995ny}, whose uncertainty dominates the SM evaluation of $a_\mu$.

\begin{figure}[t]
\begin{center}
\mbox{
\includegraphics[width=12.pc]{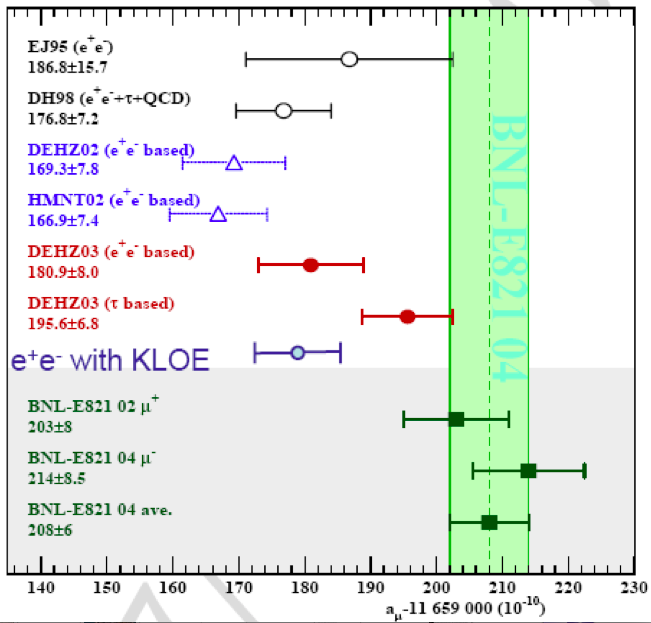}
\hspace*{0.5cm}
\includegraphics[width=10.pc]{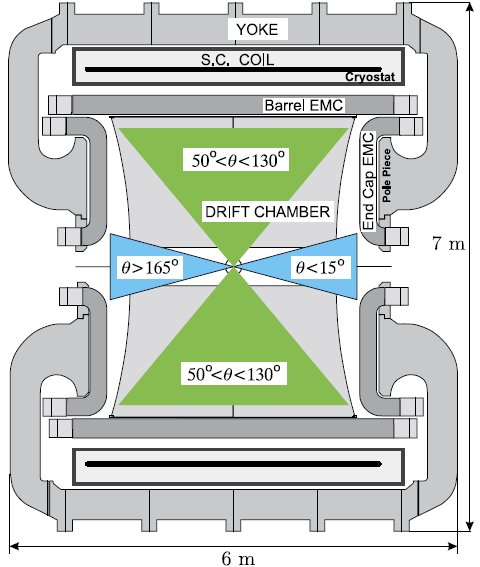}
}
\caption{Left: Comparison between different theoretical predictions of the anomalous magnetic moment of the muon  and the experimental measurement from E821 around 2004 when KLOE data were available.  From F.~Teubert presentation at ICHEP04~\cite{ichep04}. Right: Schematic view of the KLOE detector with selection regions.}
\label{fig:12} 
\end{center}
\end{figure}

At the time of KLOE05 there was a rather strong disagreement between 
$a_{\mu}^{\mathrm{HLO}}$~ value obtained using data from $\tau$ decays  into two- and four-pion final states after isospin-breaking corrections and
$e^+e^-$ based data. In particular while the SM prediction using $\tau$ data showed a difference less than $1\sigma$ with the BNL experimental result, 
the \epm\ based result disagreed by \ab3\sig\ ~\cite{Davier:2002dy}  (see Fig.~\ref{fig:12}, left). In addition
the $\pi^+\pi^-$ spectral function from $e^+e^-$ and  isospin-corrected $\tau$ data were showing a systematic deviation in shape and normalisation ~\cite{Davier:2002dy}.



KLOE05 and the following KLOE08 analyses used the so called {\it Small Angle (SA)} selection cuts: photons remaining undetected are restricted to 
a cone of $\theta_\gamma<15^\circ$ around the 
beamline (narrow cones in Fig.~\ref{fig:12}, right) and the two charged pion tracks are detected within $50^\circ<\theta_\pi<130^\circ$ (wide cones in Fig.~\ref{fig:12}, right). 
In this configuration, the direction of the photons
is reconstructed from the tracks' momenta
by closing kinematics: $\vec{p}_\gamma\simeq\vec{p}_\mathrm{miss}= -(\vec{p}_{\pi^+}
+\vec{p}_{\pi^-})$. While these cuts guarantee a high
  statistics for ISR signal events and a reduced contamination from the resonant process $e^+e^-\to
  \phi\to\pi^+\pi^-\pi^0$, in which the $\pi^0$ mimics the missing
  momentum of the photon(s), and from the final state 
radiation process $e^+e^-\to \pi^+\pi^-\gamma_\mathrm{FSR}$,  a highly 
energetic
photon emitted at small angle forces the pions also to be at small
  angles (and thus outside the selection cuts), resulting in a 
kinematic suppression of events with $M^2_{\pi\pi}< 0.35$
  GeV$^2$. The measurement of \ppg\ cross section was normalized to the DA$\Phi$NE luminosity
using large angle Bhabha scattering with 0.3\% total systematic uncertainty (using BABAYAGA), and the pion form factor was derived in the energy region between 0.35 and 0.95 GeV$^2$ 
by dividing the \ppg\ cross section by the so called radiator function obtained with NLO PHOKHARA MC generator.
Although with a limited accuracy of 1.3\% equally divided between theory and experiment, the importance of KLOE05 was twofold:
\begin{itemize}
\item It established the use of ISR as a working technique to precisely measure the hadronic cross section with a competitive accuracy of the energy scan;
\item It confirmed the ~3$\sigma$ discrepancy 
of the muon g-2 between the direct measurement and the SM prediction
based on $e^+e^-$ data, as shown in Fig~\ref{fig:12}, left. 
\end{itemize}
In addition KLOE05 allowed the KLOE Collaboration to get confidence into a novel technique which would have been refined in the following years.

\begin{figure}
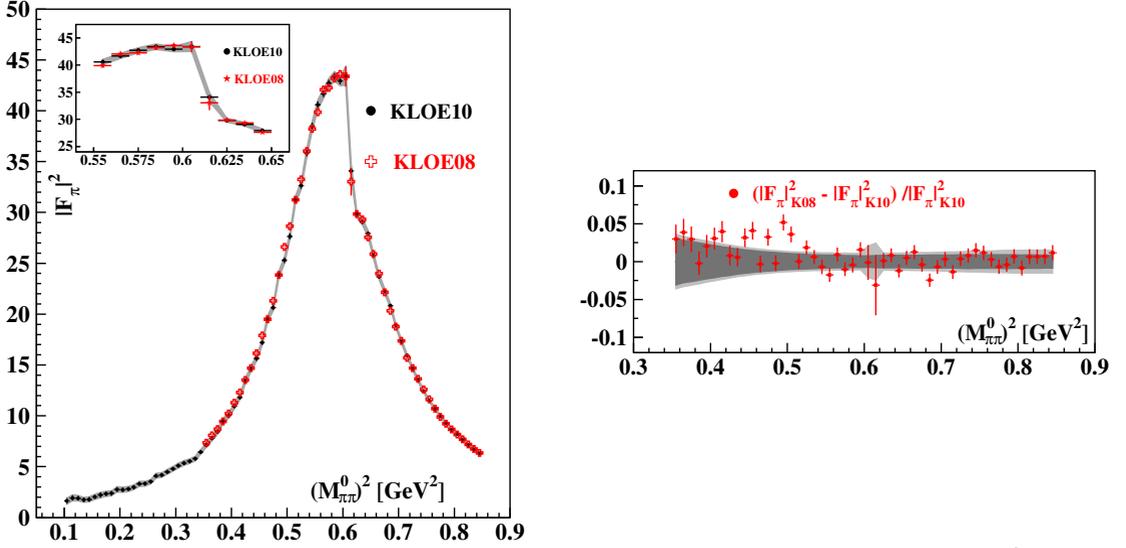

\centerline{\lfigbox fpi_kloe08_09;6.9;\kern.5cm\lfigbox dfpi_kloe08_09;7.2;}
\vglue-0.3cm
\caption{Comparison of  KLOE10  with the previous KLOE result, KLOE08. Left:  Pion form factor $|F_\pi|^2$. Right: Fractional difference between KLOE08 and KLOE10 results. The dark (light) gray band gives
  the statistical (total) error for the present result. 
Errors on KLOE08 points contain the combined statistical
  and systematic uncertainty.}
\label{fig:fpi09_08}
\end{figure}

\section{2005-2010: KLOE08 and KLOE10 analyses}
As most of the adventures in a new territory, the KLOE05 analysis was not free from errors. Particularly  an improved evaluation of the trigger correction together with a reevaluation of the luminosity (due to an improved version of Bhabha generator BABAYAGA~\cite{babayaga}), resulted in a few \% discrepancy in the two spectra below the $\rho$ peak, and a  slightly smaller value of $a^{\pi\pi}_{\mu}$ (below one standard deviation)~\cite{Ambrosino:2008aa}. 
In the meantime a new analysis called (KLOE08 in the following) was carried out with the same SA selection criteria for pion tracks and photon angles as of KLOE05 on 240 pb$^{-1}$ of data taken in 2002~\cite{Ambrosino:2008aa}.
With respect to KLOE05, KLOE08  profits  from  lower
machine  background  and  more stable  DA$\Phi$NE  operation  in  2002.
Data filters and analysis technique were improved, together with a refined
knowledge of the detector response and of the KLOE simulation and analysis,
including a new version of generator PHOKHARA,
with next-to-leading-order ISR and  FSR corrections, as well
as simultaneous emission of one ISR and one FSR photon~\cite{phokhara}.

All these refinements resulted in an improved 0.9\% total systematic uncertainty 
 consistent with KLOE05, with a total error smaller by 30\%, confirming the 3$\sigma$ discrepancy on the muon g-2.
In view of the improvements  of the analysis and the superior quality of the
2002 data, KLOE08 superseded KLOE05~\cite{Ambrosino:2008aa}.

\begin{figure}
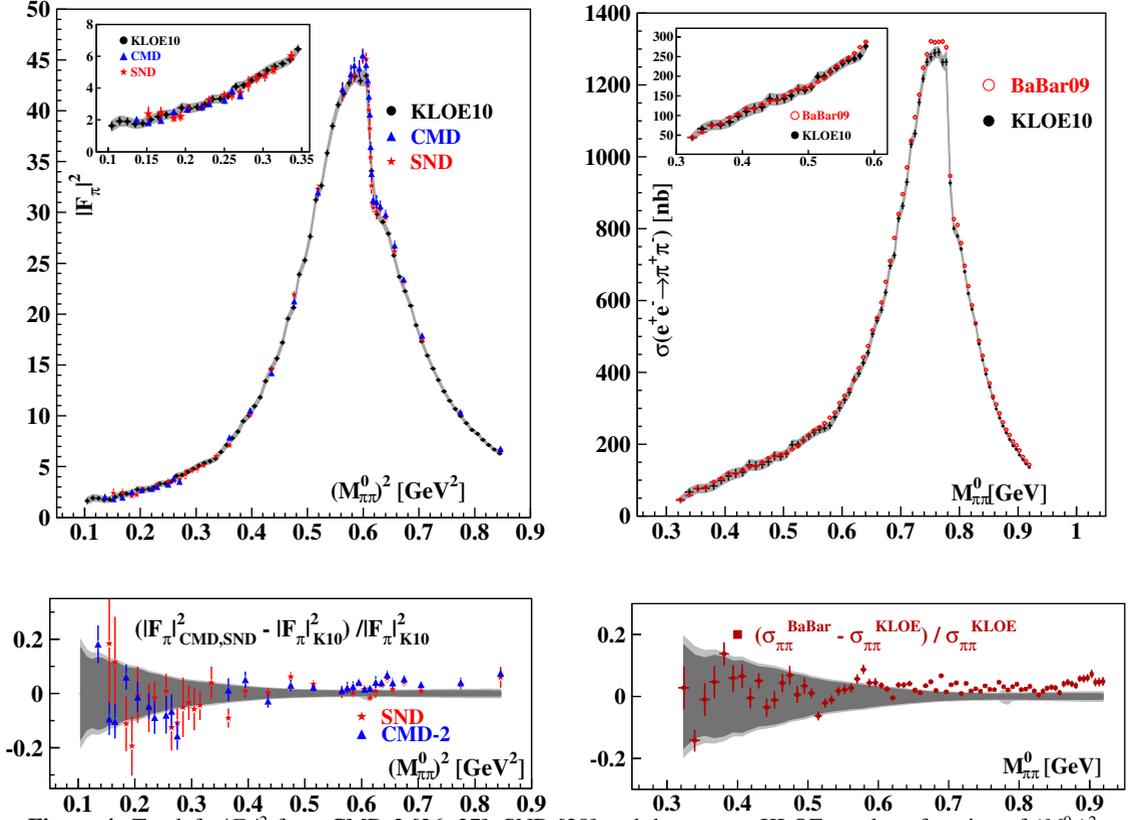

\centerline{\lfigbox fpi_kloe09_snd_cmd;6.9;\kern.5cm\lfigbox
  spp_kloe09_babar09;6.9;}
\centerline{\lfigbox dfpi_kloe09_cmd_snd;7.2;\kern.4cm\lfigbox
  diff_spp_kloe09_babar_int;7.2;}
\vglue-0.3cm
\caption{Top left: $|F_\pi|^2$ from CMD-2~\cite{CMD2, CMD2_2},
  SND~\cite{SND} and the present KLOE result as function of $(M^0_{\pi\pi})^2$. Bottom left: Fractional
  difference between CMD-2 or SND and KLOE. Top right: $\sigma^{\rm bare}_{\pi\pi}$ from
BaBar~\cite{:2009fg} and the new KLOE result as function of $M^0_{\pi\pi}$. Bottom right: Fractional
  difference between BaBar and KLOE. CMD-2, SND and BaBar data points have the total 
uncertainty attached. The
dark (light) band in the lower plots shows statistical (total) error of the KLOE result.}
\label{fig:fpi09_other}
\end{figure}

Although the KLOE08 measurement had reached a remarkable ~0.9\% systematic error, the measured region was limited to ~600 MeV, due to the SA selection requirement.
To reach the dipion threshold, a new measurement (called KLOE10 in the following) was performed where a photon was detected in the calorimeter at large polar angles, 
$50^o<\theta_{\gamma}<130^o$, in the same region where also pion tracks are detected ({\it so called} Large Angle (LA) selection)~\cite{Ambrosino:2010bv}.
However, compared to the measurements with photons at small angles, these conditions imply a reduction in statistics of about a factor of 5, and an increase of the background from the process $\phi\to\pi^+\pi^-\pi^o$, as well as the irreducible background from events with final state radiation and from $\phi$ radiative decays. To reduce the background contamination and the related uncertainties, the analysis was performed on 232.6 pb$^{-1}$ of data taken in 2006 at 1 GeV ({\it i.e.} 20 MeV below the $\phi$ mass).
The KLOE10 analysis measured $a_{\mu}^{\pi\pi}$ in the region 0.1--0.85 GeV$^{2}$ 
with a 1\% experimental systematic error and 0.9\% theoretical one dominated by uncertainty on final state radiation (FSR) model.
KLOE10 showed a good agreement with the previous KLOE
measurements (see Fig.~\ref{fig:fpi09_08}), allowing to 
cover 70\% of the leading order hadronic contribution to
the muon anomaly with ~1\% total error.
It showed also a reasonable agreement with the results from the Novosibirsk
experiments CMD-2 and SND (see Fig.~\ref{fig:fpi09_other}, left); while comparing with the BaBar result it showed
agreement below 0.4 GeV$^2$ (corresponding to a value of $M_{\pi\pi}$ of 630 MeV), while above 0.4 GeV$^2$ the BaBar result is higher by 2-3\% (Fig.~\ref{fig:fpi09_other}, right).

\begin{figure}
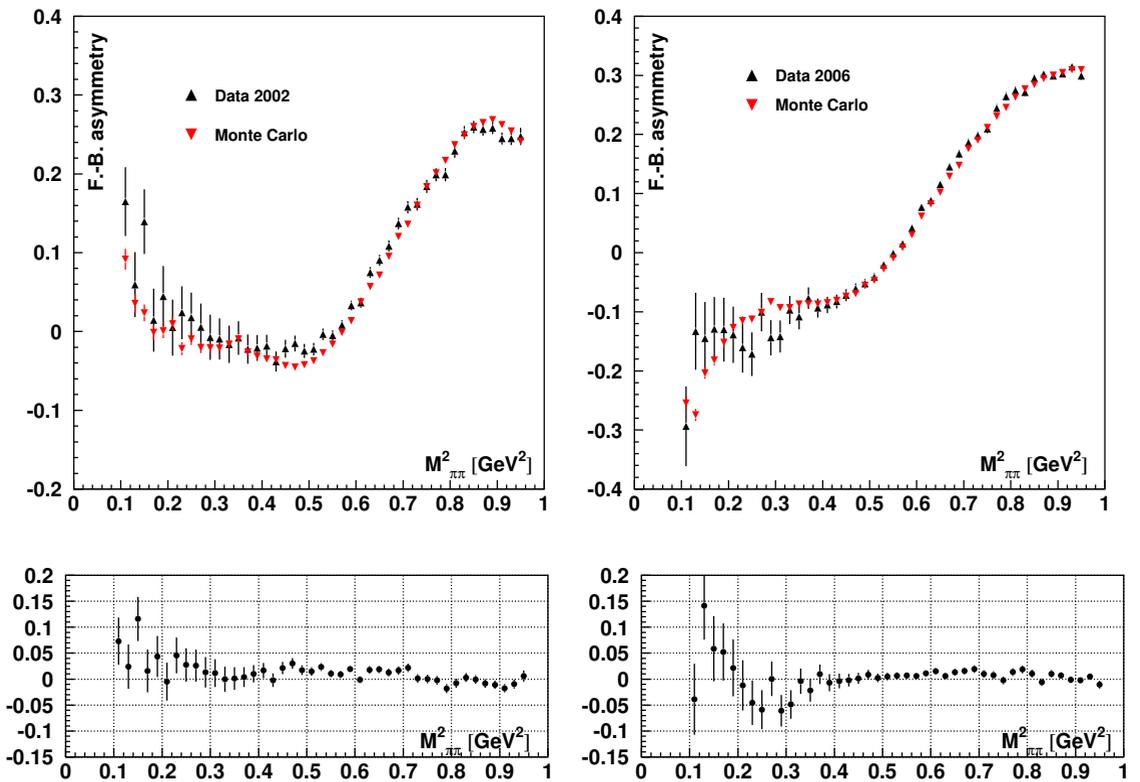

\centerline{\lfigbox asymmetry_kloe2002;6.9;\kern.5cm\lfigbox
  asymmetry_kloe2006;6.9;}
\centerline{\lfigbox diff_asymmetry_kloe2002;7.2;\kern.4cm\lfigbox
  diff_asymmetry_kloe2006;7.2;}
\vglue-0.3cm
\caption{Top left: Forward--Backward asymmetry for data taken at $\sqrt{s}=M_\phi$  in 2002, and the corresponding Monte Carlo prediction using the PHOKHARA v6.1 generator. Bottom left: Absolute difference between the asymmetries from data and Monte Carlo prediction. Top right: Forward--Backward asymmetry for data taken at $\sqrt{s} \simeq 1000$ MeV 
in 2006, and the corresponding Monte Carlo prediction using the PHOKHARA v6.1 generator. Bottom right: Absolute difference between the asymmetries from data and Monte Carlo prediction. From Ref.~\cite{actis}}
\label{fig:radret_kloeasy1}
\end{figure}

The KLOE10 data allowed also to test the validity of different models for the $\pi^+\pi^-$ final state photon emission in the forward-backward asymmetry, showing good agreement with the point-like pions ({\it so called} scalar QED) assumption (see Fig.~\ref{fig:radret_kloeasy1})~\cite{actis,Czyz:2004nq,Ambrosino:2005wk,asy}.

\section{2010-2012: KLOE12 analysis}
\begin{figure}[htb]
\begin{center}
\mbox{
\includegraphics[width=15.pc]{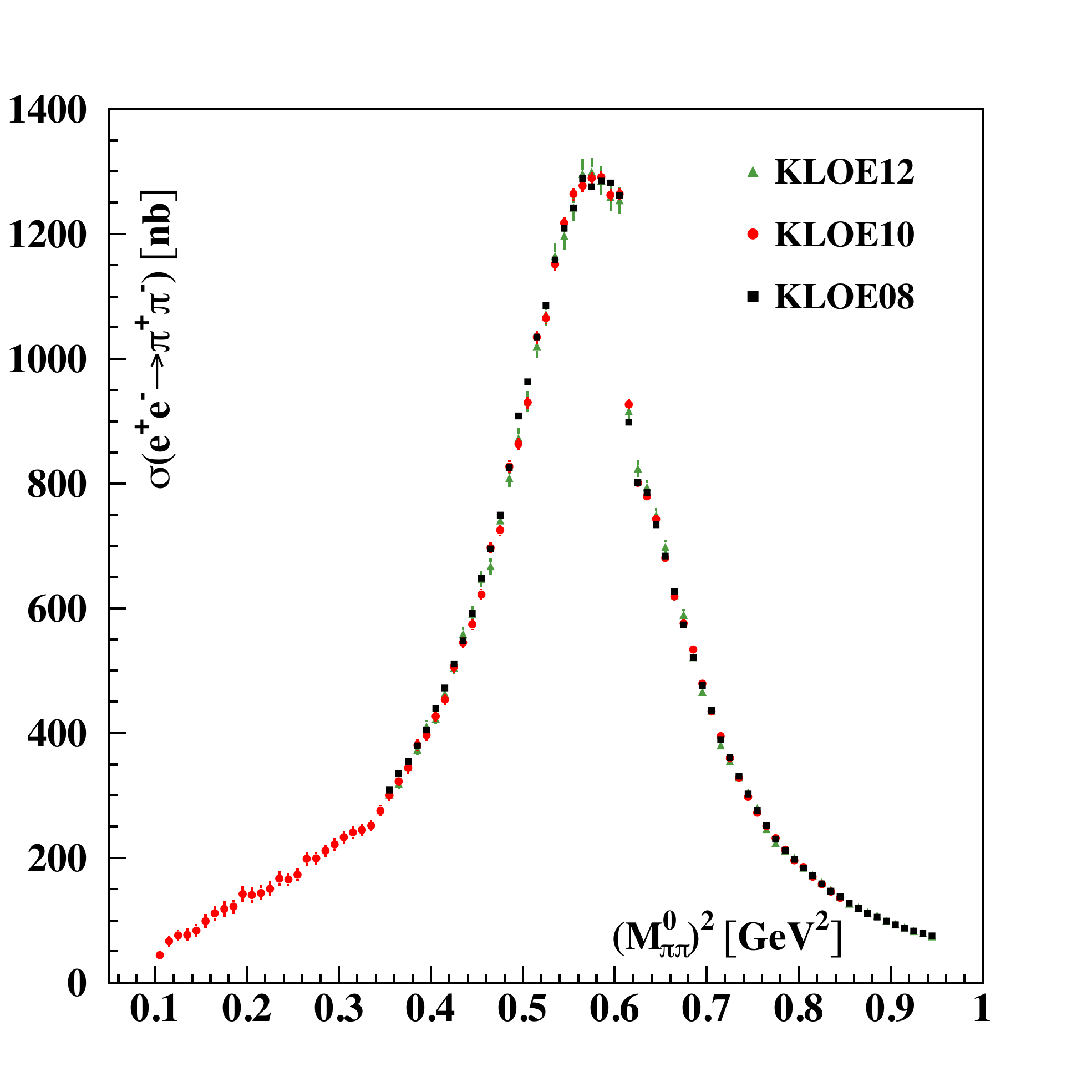}
\hspace*{0.5cm}
\includegraphics[width=15.pc]{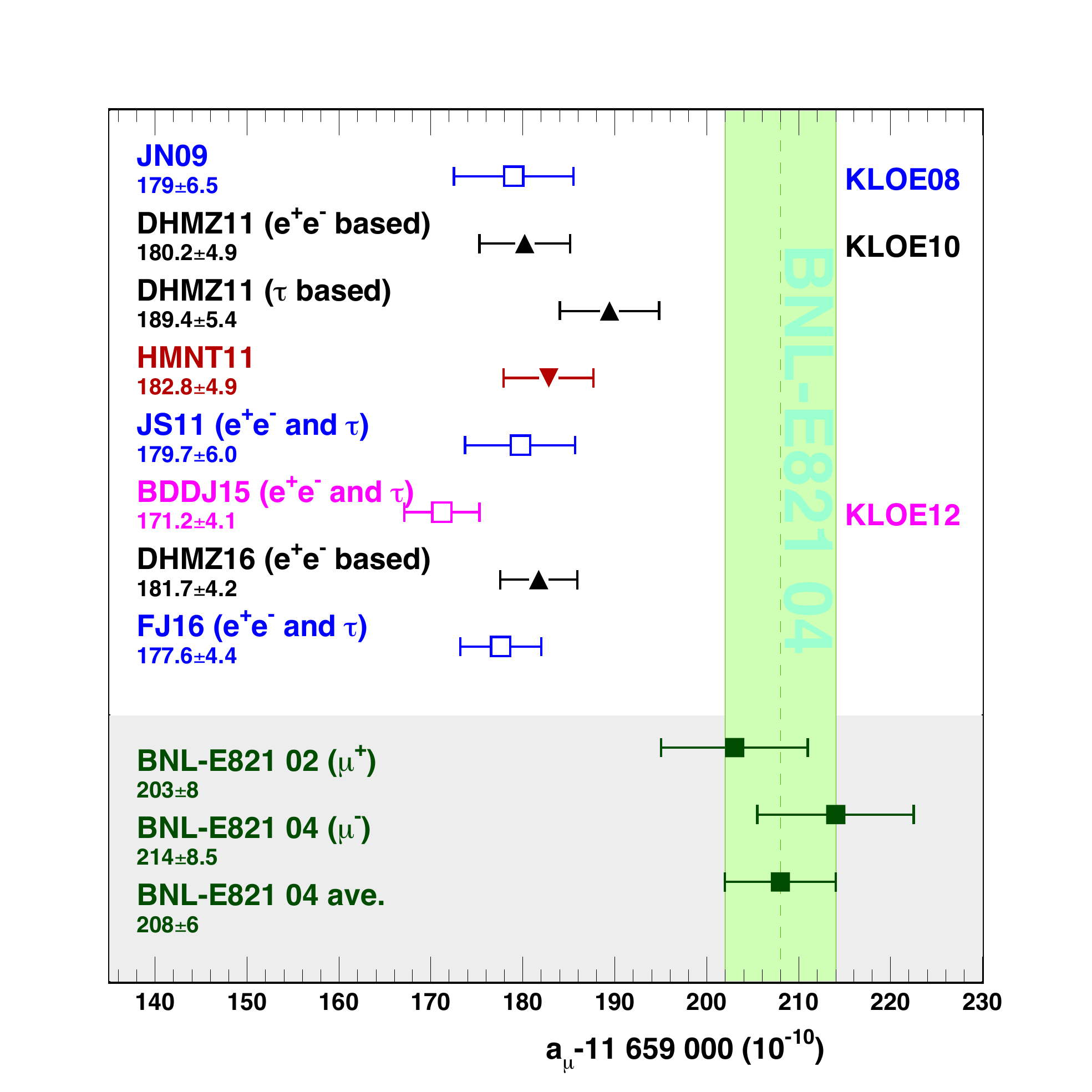}
}
\caption{Left: pion form factor $|F_\pi|^2$ obtained by \ppg/\mmg\ ratio (KLOE12) compared with previous KLOE measurements. Right: compilation of recent results for $ a_\mu$ compared with the experimental value. The SM predictions refer to: 
JN09~\cite{Jegerlehner:2009ry}, DHMZ11~\cite{Davier:2010nc}, HLMNT11~\cite{Hagiwara:2011af}, JS11~\cite{Jegerlehner:2011ti}, BDDJ15~\cite{Benayoun:2015gxa}, DHMZ16~\cite{Davier:2016iru}, FJ16~\cite{Jegerlehner:2017lbd}.
KLOE08~\cite{Ambrosino:2008aa}, KLOE10~\cite{Ambrosino:2010bv}, and KLOE12~\cite{Babusci:2012rp} refer to the compilations where KLOE data are included.}
\label{fig:2} 
\end{center}
\end{figure}

While KLOE08 and KLOE10 were independent measurements with respect to the 
 data set and angular selection, they were 
both normalized to the DA$\Phi$NE luminosity, and used the radiation function to obtain the pion form factor and $a_{\mu}$. 
As shown in Ref.~\cite{:2009fg} it's possible to extract the pion form factor by normalizing to the muon ISR differential cross section.
In this approach 
the integrated luminosity as well as the radiation function H cancel in the ratio, as does the vacuum polarisation. In addition using the same fiducial volume, acceptance corrections to \ppg\ and \mmg\ spectra almost cancel resulting in a per mille systematic uncertainty.
However the price to pay is to perform an additional analysis on muon events at subpercent level.
The same sample of 239.2 pb$^{-1}$ of KLOE08 was
analyzed with the small angle photon selection (called KLOE12 in the following)~\cite{Babusci:2012rp}. 
While the analysis for \ppg\ is essentially the same as for KLOE08, the analysis for \mmg\ is new and is based on the following main features: 1) separation between \mmg\ and \ppg\ 
events obtained assuming the final state of two charged particles with equal track mass \mtrk\ and one photon: the \mtrk\ $<$ 115
MeV (\mtrk\ $>$ 130 MeV) selection leads to $9\times 10^5$ ($3.1\times 10^6$) candidate
\mmg\ (\ppg) events, this selection is checked
against other techniques, such as a kinematic fit or tighter cuts on the quality of the charged tracks, all providing
consistent results; 2) trigger, particle identification and tracking efficiencies checked from data control samples. 
The \mmg\ cross section measurement  was compared with the one obtained by PHOKHARA MC~\cite{phokhara}, and  a good agreement is
found~\cite{Babusci:2012rp}. Then the pion form factor has been extracted and compared with the one from KLOE10, showing good agreement (see  Fig.~\ref{fig:2}, left).
The value for $a_\mu^{\pi\pi}$ computed with KLOE12 was in good agreement with the previous KLOE results, providing therefore a strong cross check of the systematics and confirming the  3$\sigma$ discrepancy between the experimental value and the Standard Model prediction of the muon g-2 (see  Fig.~\ref{fig:2} right).

\section{2013-2016: Measurement  of the running of the fine structure constant below 1 GeV}
With about 2 fb$^{-1}$ of luminosity collected in 2004-2005, the KLOE-2 Collaboration - the successor of KLOE - embarked in a full statistics analysis of the \mmg\ cross section  with a twofold motivation:
\begin{enumerate}
\item  to search for a light vector boson~\cite{Curciarello:2016jbz};
\item to measure  the  running  of  the  $QED$ coupling constant $\alpha(s)$
below 1 GeV.
\end{enumerate}
The analysis of the so called $U-$boson is in progress. 
In the following we will discuss 
the measurement of the running of 
the fine structure constant $\alpha$
in the time-like region 0.6$<\sqrt{s}<$0.975 GeV~\cite{KLOE-2:2016mgi}. \\

\begin{figure}[h!]
\begin{center}
\mbox{
\includegraphics[width=16.pc]{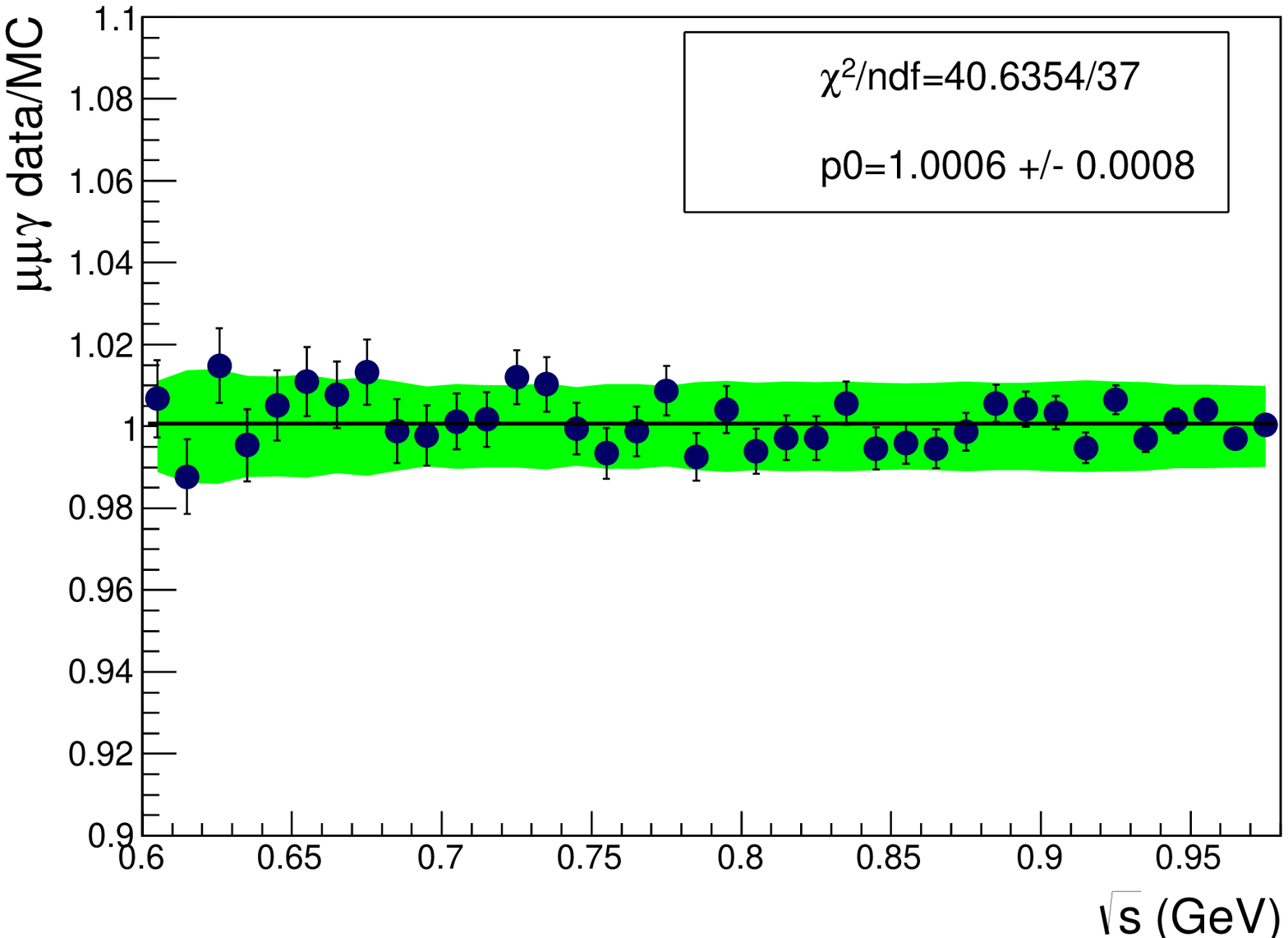}
\hspace*{0.5cm}
\includegraphics[width=16.pc]{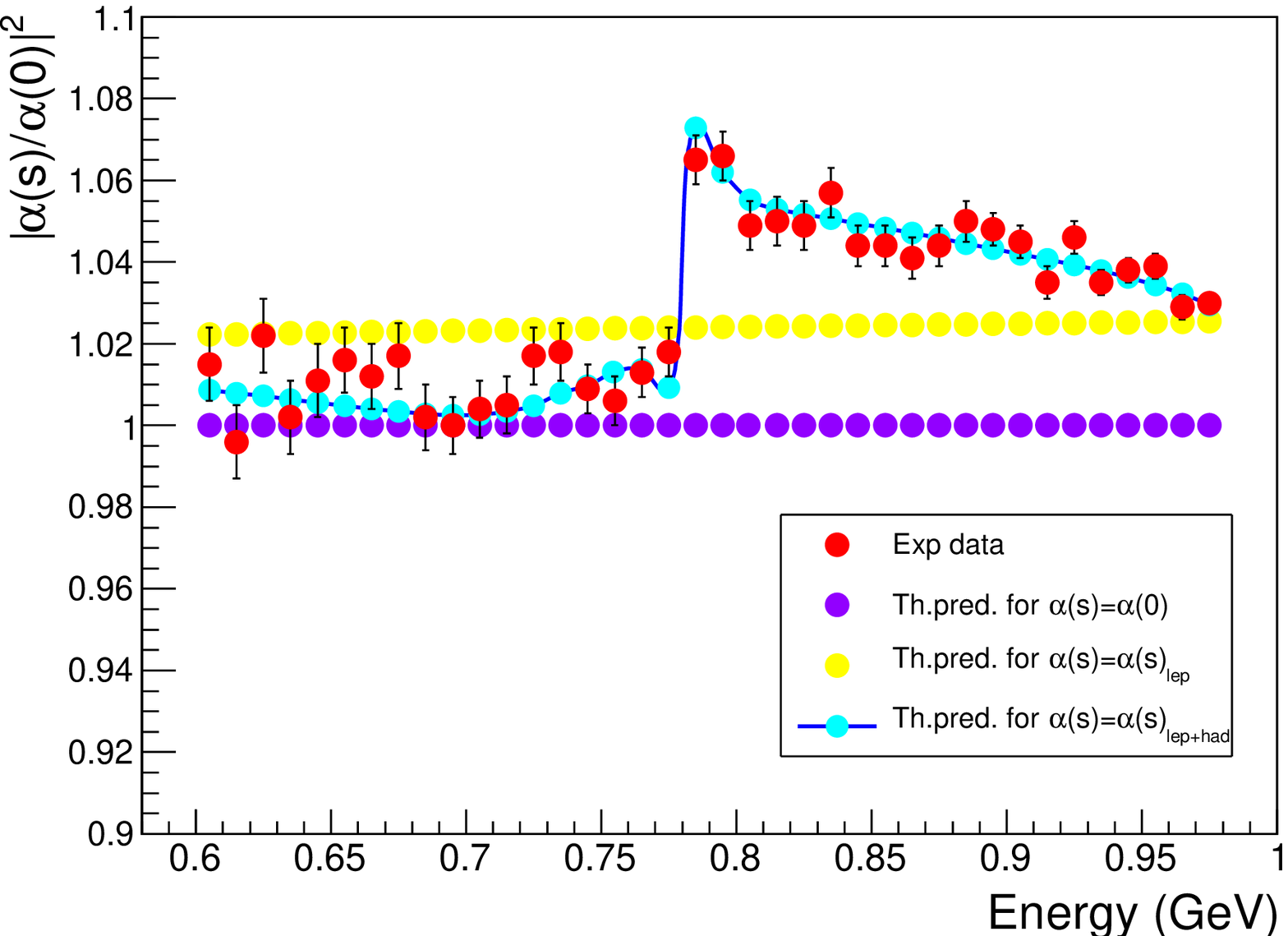}
}
\caption{Left: Ratio between 
the measured differential $\mu^+\mu^-\gamma$ cross section 
and the MC prediction from PHOKHARA. The green band shows the systematic error. Right:
The square of the modulus of the running $\alpha(s)$ in units of $\alpha(0)$ compared with the 
prediction (provided by the {\tt alphaQED} package \cite{fj}) 
as a function of the dimuon invariant mass. The red points are the 
KLOE data with statistical errors; the violet points are the theoretical prediction for a fixed coupling ($\alpha(s)$ = $\alpha(0)$);
the yellow points are the prediction with only virtual lepton pairs contributing to the shift $\Delta\alpha(s)$ = $\Delta\alpha(s)_{lep}$, 
and finally the points with the solid line are the full QED prediction with both lepton and quark pairs contributing to the shift $\Delta\alpha(s)$ = $\Delta\alpha(s)_{lep+had}$.}
\label{mmg_abs_q.eps}
\end{center}
\end{figure}

The strength of the coupling constant is measured as a function of the momentum transfer of the
exchanged photon $\sqrt{s}=M_{\mu\mu}$ where $M_{\mu\mu}$ is the  $\mu^+\mu^-$ invariant mass.
\,The value of $\alpha(s)$ is extracted from the ratio of the  differential cross section 
 for the process $e^+e^- \rightarrow \mu^+\mu^- \gamma(\gamma)$ with the photon emitted in the Initial State (ISR) to the corresponding cross section obtained from Monte Carlo (MC) simulation 
with the coupling set to the constant value $\alpha (s)= \alpha(0)$:
\begin{equation}
\lvert \frac{\alpha(s)}{\alpha(0)}\rvert^2= \frac{d\sigma_{data} (e^+e^- \rightarrow \mu^+\mu^- \gamma(\gamma))\vert_{ISR}/d\sqrt{s}}{d\sigma^{0}_{MC}(e^+e^- \rightarrow \mu^+\mu^- \gamma(\gamma))\vert_{ISR}/d\sqrt{s}}
\label{our_method}
\end{equation}
To obtain the ISR cross section, the observed cross section must be corrected for events with one or more photons in the final state (FSR). This has been done by using the PHOKHARA MC event generator,  which includes  next-to-leading-order ISR and FSR contributions~\cite{phokhara}.
Figure~\ref{mmg_abs_q.eps}, left, shows the ratio of the $\mu^+\mu^-\gamma$ cross-section from data with the corresponding NLO QED calculation from PHOKHARA generator including the Vacuum Polarization effects.

The agreement between the two cross sections is excellent. 

We use  Eq.~(\ref{our_method}) 
to extract the running of the effective QED coupling constant $\alpha(s)$. 
By setting in the MC the electromagnetic coupling to the constant value $\alpha(s)$ = $\alpha(0)$, the hadronic contribution to the photon propagator,
 with its
characteristic
 $\rho-\omega$ interference structure, is clearly visible, see Fig.~\ref{mmg_abs_q.eps}, right. The prediction from Ref.\cite{fj}  is also shown. While the leptonic part is obtained by perturbation theory, the hadronic contribution
 to $\alpha(s)$ is
obtained via an evaluation in terms of a weighted average
compilation of $R_{had}(s)$, based on the available experimental $e^+e^- \to
\mathrm{hadrons}$ annihilation data (for an up to date compilation see~\cite{Jegerlehner:2015stw} and references therein). 


For comparison, the prediction with constant coupling ({\it no running}) and with only lepton pairs contributing to the running of $\alpha(s)$ is given.\\
By including statistical and systematics errors, we exclude the only-leptonic hypothesis at $6\,\sigma$
which is the strongest direct evidence ever achieved by a collider 
experiment~\footnote{The first evidence for the hadronic VP came from the ACO experiment which found an evidence at $3\sigma$ of the $\phi$ contribution to the process $e^+e^-\to\mu^+\mu^-$ in the  region $\pm 6$ MeV around the $\phi$ peak (at 1019.4 MeV)~\cite{Augustin:1973ep}. The strongest evidence for hadronic VP effect comes from the muon g-2 experiment; at CERN it was determined at more than $7\sigma$ level~\cite{Bailey:1977mm,Bailey:1978mn}.}.
\subsection{Extraction of Real and Imaginary part of $\Delta\alpha(s)$}
By using the definition of the running of $\alpha$ 
the real part of the shift $\Delta\alpha(s)$ can be expressed in terms of its imaginary part and $|\alpha(s)/\alpha(0)|^2$:
\begin{equation}
\rm Re\,\Delta \alpha = 1-\sqrt{\vert \alpha(0)/\alpha(s) \vert^2-(\rm Im\,\Delta \alpha)^2}.
\label{re_delta_alpha}
\end{equation}  
The imaginary part of $\Delta\alpha(s)$ can be related to the 
total cross section $\sigma(e^+e^-\to\gamma^*\to anything)$,
  where the precise relation reads~\cite{Eidelman:1995ny,Jegerlehner:2009ry,fred}:
 \vspace{0.5cm}
${\rm Im}\,\Delta \alpha = - \frac{\alpha}{3}\,R(s)$,
with $R(s) = \sigma_{tot}/\frac{4\pi\vert\alpha(s)\vert^2}{3s}$.
$R(s)$ takes into account leptonic and hadronic contribution
$R(s)=R_{lep}(s)+R_{had}(s)$, 
where the leptonic part corresponds to the production
of a lepton pair at lowest order taking into account mass effects:

\begin{equation}
R_{lep}(s)=\sqrt{1-\frac{4m_l^2}{s}} \left(1+\frac{2m_l^2}{s}\right), \;\; (l=e,\mu,\tau). 
\end{equation}

In the energy region around the $\rho$-meson we can approximate the hadronic cross section by the 2$\pi$ dominant contribution:

\begin{equation}
R_{had}(s)= \frac{1}{4} \left(1-\frac{4m_\pi^2}{s} \right)^\frac{3}{2} \vert F_\pi^0(s) \vert ^2,
\end{equation}

where $F_\pi^0$ is the pion form factor deconvolved:
$\lvert F_\pi^0(s)\rvert ^2 = \lvert F_\pi(s) \rvert ^2\left\lvert \frac{\alpha(0)}{\alpha(s)}\right\rvert^2$.

\vspace{0.5cm}

\begin{figure}[htp!]
\begin{center}
\mbox{
\includegraphics[width=16.pc]{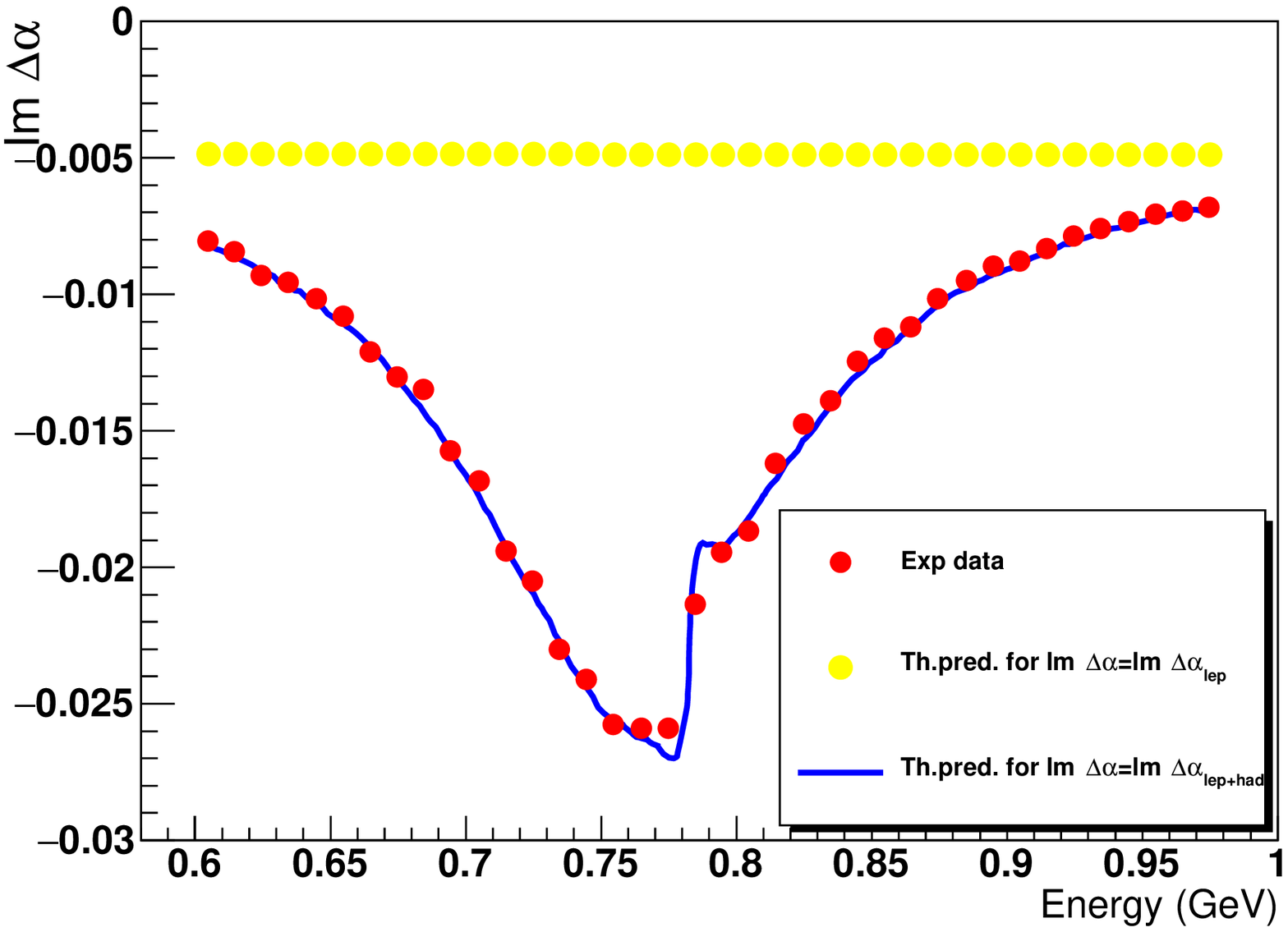}
\hspace*{0.5cm}
\includegraphics[width=16.pc]{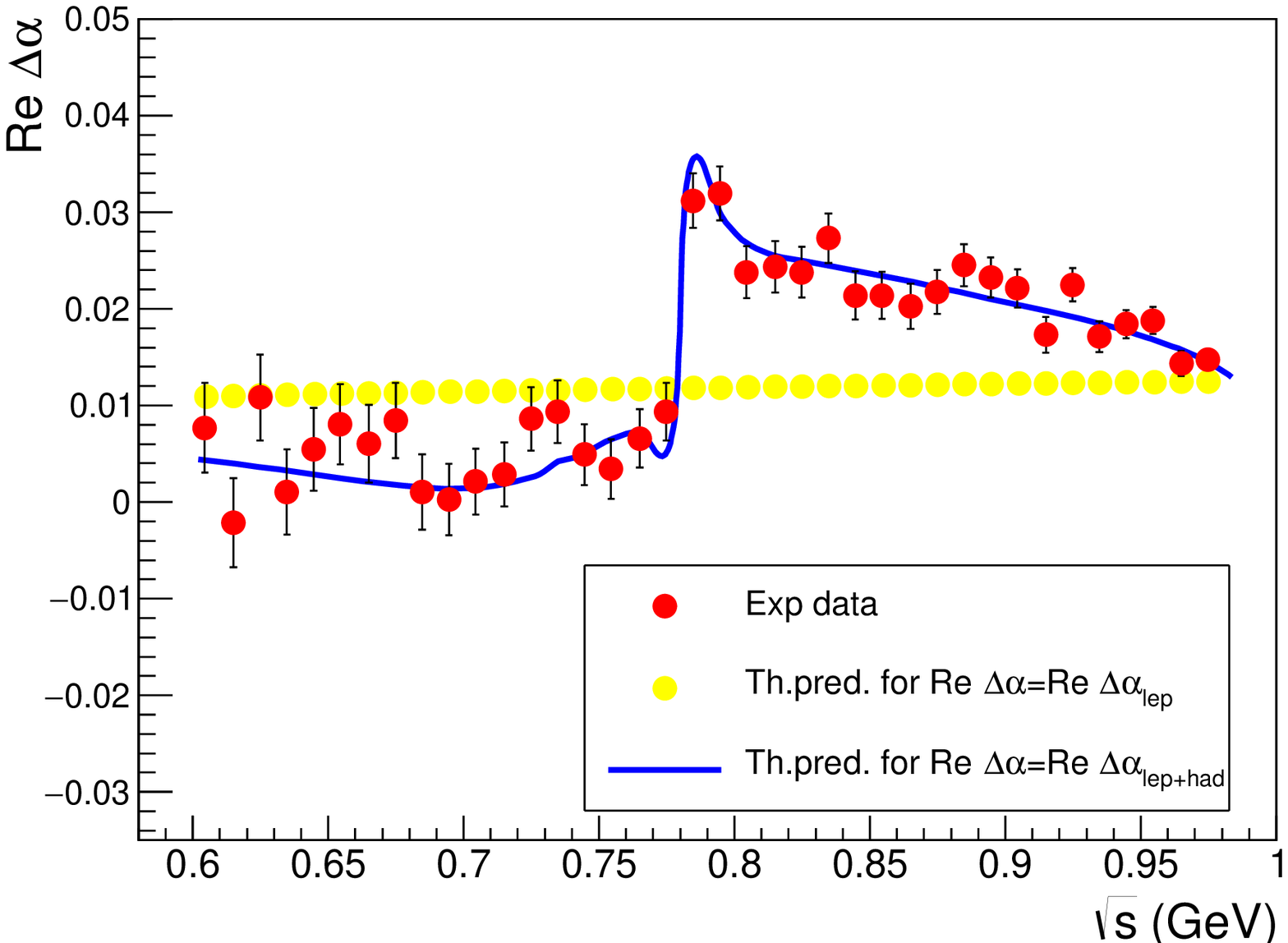}
}
\caption{Left: $\rm Im\,\Delta \alpha$ extracted from the KLOE data 
compared with the values provided by {\tt alphaQED} routine (without the KLOE data) for
 $\rm Im\,\Delta \alpha=\rm Im\,\Delta \alpha_{lep}$ (yellow points) and $\rm Im\,\Delta \alpha=\rm Im\,\Delta \alpha_{lep+had}$ only for $\pi\pi$ channels (blue solid line). Right:
$\rm Re\,\Delta \alpha$ extracted from the experimental data with only the statistical error included compared with the {\tt alphaQED} prediction (without the KLOE data) 
 when $\rm Re\,\Delta \alpha=\rm Re\,\Delta \alpha_{lep}$ (yellow points) and $\rm Re\,\Delta \alpha=\rm Re\,\Delta \alpha_{lep+had}$ (blue solid line). 
 }
\label{im_delta_alpha}
\end{center}
\end{figure}

The results obtained for the $2\pi$ contribution to the imaginary part of $\Delta\alpha(s)$ by using the KLOE pion form factor measurement\cite{Babusci:2012rp},
are shown in Fig.~\ref{im_delta_alpha} and
compared 
with the values given by the $R_{had}(s)$ compilation of Ref.~\cite{fj} using only the $2\pi$ channel, with the KLOE data removed (to avoid correlations).  

The extraction of the $\rm Re\,\Delta \alpha$ has been performed using the Eq.~(\ref{re_delta_alpha})
and it is shown in Fig.~\ref{im_delta_alpha}, right. The experimental data with only the statistical error included have been compared 
with the {\tt alphaQED} prediction when $\rm Re\,\Delta \alpha=\rm Re\,\Delta \alpha_{lep}$ (yellow points in the colour Figure) and $\rm Re\,\Delta \alpha=\rm Re\,\Delta \alpha_{lep+had}$ (dots with solid line). 
As can be seen, an excellent agreement for $\rm Re\,\Delta\alpha(s)$ has been obtained with the data-based compilation.



\clearpage
Finally $\rm Re\,\Delta \alpha$ has been fitted by a sum of the leptonic and hadronic contributions, where
the hadronic contribution is parametrized as a sum of the $\rho(770)$, $\omega(782)$ and $\phi(1020)$ resonance components and a non-resonant term. 
The product of the branching fractions has been extracted:
\begin{equation}
BR(\omega\to\mu^+\mu^-)BR(\omega\to e^+e^-) = (4.3\pm1.8_{stat}\pm2.2_{syst})\cdot 10^{-9},
\end{equation}
where the first error is statistical and the second systematic. 
By multiplying by the phase space factor 
$\xi =\Big(1+2\frac{m^2_\mu}{m^2_{\omega}}\Big)\Big(1-4\frac{m^2_\mu}{m^2_{\omega}}\Big)^{1/2}$
 and assuming lepton universality, $BR(\omega\to\mu^+\mu^-)$ can be extracted:
\begin{equation}
BR(\omega\to \mu^+\mu^-)=(6.6\pm1.4_{stat}\pm1.7_{syst})\cdot 10^{-5} 
\end{equation}
compared to $BR(\omega\to\mu^+\mu^-) =(9.0\pm3.1)\cdot 10^{-5}$ from PDG~\cite{pdg}.\\

\section{Conclusions and Outlook}

The last 20 years have seen a flourishing use of the ISR to measure the hadronic cross sections at flavor factories. Being more than a poor man’s
alternative to scan energies the ISR represents an alternate, independent and complementary way to determine those cross sections with different systematic errors. Such a success story was possible thanks to dedicated work both from the 
theoreticians developing Monte Carlo generators like PHOKHARA and BABAYAGA@NLO which were fundamental to achieve sub percent precision, as well as from experimental groups using ISR not only to determine hadronic cross sections but also for spectroscopy studies. In that respect the KLOE measurements of hadronic cross section 
was a significant contribution to this field~\footnote{As underlined also by more than 600 citations of the corresponding articles~\cite{Aloisio:2004bu,Ambrosino:2008aa,Ambrosino:2010bv,Babusci:2012rp}.}:
\begin{itemize}
\item It proved the ISR as a working tool to precisely measure the hadronic cross sections;
\item 
It confirmed the $3\sigma$ discrepancy between the Standard  Model prediction for the g-2 and 
the measured value, and the disagreement with the SM prediction based on $\tau$ data. These two sets of data are now in agreement with each other, once the vector meson and $\rho-\gamma$ mixing effects are properly accounted 
for~\cite{rhomix};

\item It allowed a first direct measurement of the time-like  complex  running
$\alpha(s)$. That procedure would be helpful to solve a problem in the treatment of the narrow resonances to $a_{\mu}$ and to the running of $\alpha$~\cite{Jegerlehner:2015stw};
\item By the close contact and interactions with theorists it contributed to the inter-comparisons and refinements of the Monte Carlo codes~\cite{actis}. 
\end{itemize}
Thanks to the worldwide (experimental and theoretical) efforts the accuracy of
leading-order hadronic vacuum polarization contribution, $a_{\mu}^{\mathrm{HLO}}$, has improved to $\sim 4\times 10^{-10}$. Although corresponding to a remarkable fractional uncertainty of 0.6\%, it still 
constitutes the main uncertainty of the SM prediction to $a_\mu$.
 New experiments at Fermilab and J-PARC aiming at a  fourfold improved precision are underway~\cite{grange,saito}.
Together with a fourfold improved precision on the experimental side, 
an improvement of the LO hadronic contribution is highly desirable.
An interesting possibility in this respect is a novel
approach aiming to determine $a_{\mu}^{\mathrm{HLO}}$ 
from a direct space-like measurement of $\alpha(t)$ as recently proposed in Refs.~\cite{Abbiendi:2016xup,Calame:2015fva} (see also~\cite{trentadue}).
\section{Acknowledgements}
The ISR measurement at KLOE was first of all a story of passion, friendship, committement and dedication. 
I would like to express my gratitude to the 
members of the KLOE/KLOE-2 Collaboration who made this possible and with whom I had the privilige to collaborate, particularly: P. Beltrame, F. Curciarello, V. De Leo, A. Denig, S. Di Falco, M. Incagli, W. Kluge, J. Lee-Franzini, P. Lukin, G. Mandaglio, S. M\"{u}ller, F. Nguyen, A. Palladino, and B. Valeriani. Many friends and colleagues gave useful suggestions, help, criticism and support during the development of the analyses:
A. Arbuzov, M. Benayoun, C.M. Carloni Calame, H. Czy\.z, S. Eidelman, G. Fedotovich, F. Ignatov, S. Ivashyn, F. Jegerlehner, A. Keshavarzi, J.H. K\"{u}hn, G. Montagna, O. Nicrosini, A. Nyffeler, L. Pancheri, M. Passera, F. Piccinini, G. Rodrigo, O. Shekhtostova, T. Teubner, and L. Trentadue.
To all of them, and to the whole Radio Monte Carlo Working Group, goes my sincere gratitude and appreciation.
Finally I would like to thank 
Rita Bertelli for her precious work and assistance during all these years.



The KLOE-2 Collaboration warmly thanks the former KLOE colleagues for the access to the data collected during the KLOE data taking campaign.
We thank the DA$\Phi$NE team for their efforts in maintaining low background running conditions and their collaboration during all data taking. We want to thank our technical staff: 
G.F. Fortugno and F. Sborzacchi for their dedication in ensuring efficient operation of the KLOE computing facilities; 
M. Anelli for his continuous attention to the gas system and detector safety; 
A. Balla, M. Gatta, G. Corradi and G. Papalino for electronics maintenance; 
C. Piscitelli for his help during major maintenance periods. 
This work was supported in part 
by the Polish National Science Centre through the Grants No.\
2013/08/M/ST2/00323,
2013/11/B/ST2/04245,
2014/14/E/ST2/00262,
2014/12/S/ST2/00459.

%

\end{document}
